\documentclass[conference]{vldb}  

\usepackage{graphics} 
\usepackage{epsfig} 
\usepackage{mathptmx} 
\usepackage{times} 
\usepackage{amsmath} 
\usepackage{amssymb}  
\usepackage{makecell}
\usepackage{xcolor}
\usepackage{subfig}
\usepackage{url}
\usepackage{listings}
\usepackage{comment}
\usepackage{xspace}
\newcommand{\Soteria}[0]{Soteria\xspace}

\vldbTitle{Soteria: A Provably Compliant User Right Manager Using a Novel Two-Layer Blockchain Technology}
\vldbAuthors{Stanford University$^{1}$, HTC DeepQ$^{2}$, and NTU$^{3}$%
}
\vldbDOI{https://doi.org/10.14778/xxxxxxx.xxxxxxx}
\vldbVolume{12}
\vldbNumber{xxx}
\vldbYear{2020}

\begin{document}

\title{\Soteria: A Provably Compliant User Right Manager \\ Using a Novel Two-Layer Blockchain Technology \\
}
\author{ 
 Wei-Kang Fu$^{3}$ \quad  Yi-Shan Lin$^{2}$ \quad Giovanni Campagna$^{1}$ \\
De-Yi Tsai$^{3}$ \quad Chun-Ting Liu$^{2}$ \quad Chung-Huan Mei$^{2}$ \\
Edward Y. Chang$^{1,2}$ \quad Monica S. Lam$^{1}$  \quad Shih-Wei Liao$^{3}$ \\ 
Stanford University$^{1}$ \quad HTC DeepQ$^{2}$ \quad National Taiwan University$^{3}$ \\
echang@cs.stanford.edu \quad lam@cs.stanford.edu
}
\maketitle
\thispagestyle{empty}
\pagestyle{empty}
\sloppy

\begin{abstract}
Soteria is a user right management system designed to safeguard user-data privacy in a transparent and provable manner in compliance to regulations such as GDPR and CCPA. Soteria represents user data rights as formal executable sharing agreements, which can automatically be translated into a human readable form and enforced as data are queried.  To support revocation and to prove compliance, an indelible, audited trail of the hash of data access and sharing agreements are stored on a two-layer distributed ledger. The main chain ensures partition tolerance and availability (PA) properties while side chains ensure consistency and availability (CA), thus providing the three properties of the CAP (consistency, availability, and partition tolerance) theorem. Besides depicting the two-layer architecture of Soteria, this paper evaluates representative consensus protocols and reports performance statistics. 
\end{abstract}

\section{Introduction}
\label{sec:intro}

Artificial intelligence (AI) has the potential to improve the quality of many application domains. In order to effectively train AI models, an application typically requires large quantities of personal information (PI) from users. To address data privacy issues, regulations such as GDPR~\cite{EUdataregulations2018} 
and CCPA~\cite{ccpabyCA20192020} in the general domain and HIPAA in the medical domain need to be upheld. 
The eight rights of GDPR and the six of CCPA can be summarized into
three categories: 1) right to {\em consent}, rectify, and delete; 2)
right to be {\em informed}; and 3) right to {\em access} and transfer.
\begin{itemize}
    \item Consent: users must explicitly opt-in and always have the ability to opt-out of PI collection.
    \item Informed: users have the right to know how their PI is collected, used, and shared.
    \item Access: users have the right to access his/her own data, transfer data to another 
    person or entity, and erase data.
\end{itemize}

Businesses are required to comply with the {\em consumer rights} in a {\em provable} way. A robust privacy-preserving PI governance platform is needed to 
ensure that all transactions (including permission, revocation,
data access, and deletion) on PI leave indelible and consistent
records for public audit.
Provability is 
essential in the court of law to resolve he-said-she-said cases.  


To protect privacy rights in a provable manner, we propose 
\Soteria, a user-right management system with a distributed ledger platform. 
The user-right management system provides a formal end-to-end solution that automatically maps user agreements in natural language into formal compliance code.  Our {\em executable sharing agreements} (ESA) are a formal representation of sharing agreements that can specify a superset of the rights protected under GDPR and CCPA.  These agreements can be translated into plain language automatically so users can understand them; they are translated into formal Satisfiability Modulo Theory (SMT) formulas for enforcement. 
We use a distributed ledger to support auditability and revocability.  We create an indelible trail of records by creating a log of every agreement signed and every query made, and a hash of the log is stored on the distributed ledger.

\Soteria ensures performance scalability in terms of both latency and throughput.
\Soteria employs
a multi-layer blockchain architecture (we previously
proposed in \cite{8613656}) to allow all three CAP (consistency, 
availability, and partition tolerance) properties \cite{EB-cap-theorem} to be 
simultaneously satisfied. 
The base-layer blockchain guarantees PA (partition tolerance and availability)
properties to ensure scalability, whereas
the side chains guarantee CA (consistency and availability) properties
to ensures provability.  We detail our protocol selections for the base 
blockchain and side chains
in Section~\ref{sec:protocols}.

The rest of this paper is organized into four sections.
Section~\ref{sec:architecture} defines
depicts \Soteria's architecture and its user right management system.  
Section~\ref{sec:protocols} presents and qualitatively 
evaluates four representative consensus protocols 
that satisfy consistency and availability (CA) requirements of \Soteria's
side chains. In Section~\ref{sec:experiments}, we report our quantitative performance 
evaluation on Tendermint and Stellar protocols and justify our selection of 
Tendermint. We offer our concluding 
remarks in Section~\ref{sec:con}.

\section{Architecture of Soteria}
\label{sec:architecture}

We use the terms {\em user} and {\em consumer} interchangeably to refer to the owner of data.
(Data consumer is the term defined in CCPA to refer to an application user whose data
the regulation aims to protect.)
We use {\em business} and {\em company} to refer to the collector and custodian of user/consumer data.
A user, and a business with that user's consent, can access 
the data collected from the user stored at the business.

\begin{figure*}[ht!]
\vspace{-.1in}
\begin{center}
\includegraphics[width=16cm,height=8.5cm]{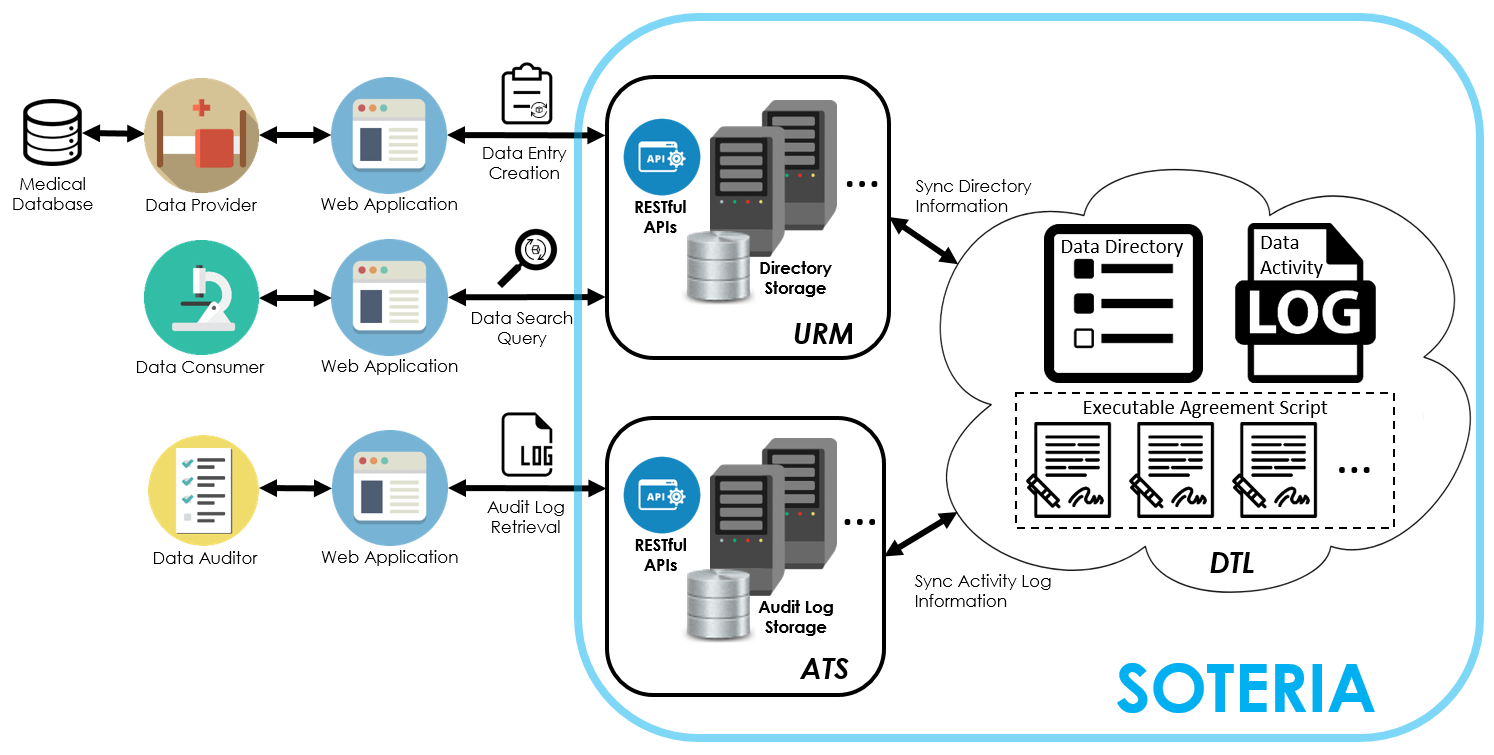}
\end{center}
\vspace{-.3in}
\caption{\Soteria Components: URM, DTL, and ATS.}
\label{fig:soteria-arch}
\end{figure*}

\newpage
\subsection{Components and Design Goals}

Figure~\ref{fig:soteria-arch} depicts components of \Soteria: 
URM, DTL, and ATS.
\begin{itemize}
\item 
{\em User right management} (URM): URM stores information about all user data and metadata 
that are collected and stored by a company.
\item {\em Distributed ledger} (DTL): DTL is our multi-layer blockchain that satisfies \Soteria's 
requirements including privacy, throughput and latency. 
\item {\em Audit trail service} (ATS): ATS
stores incurred transactions on data items for transparent auditing.
\end{itemize}

\Soteria is designed to address each of the following challenges: 
\begin{enumerate}
    \item Today users are asked to consent to long hard-to-read agreements. How can we write agreements that can be understandable to users?
    \item How do we ensure that the agreements users sign translate into a faithful implementation? 
    \item Consumers are not expected to keep records of all the agreements signed, how can a company prove that it is compliant?  In particular, how do we ensure that no accesses to revoked data are performed? 
\end{enumerate}

While Section~\ref{sec:protocols} depicts the the ledger design in DTL for addressing 
the provability requirement, the remainder of this section presents how URM complies
with regulations of consent, informed, and access described in Section~\ref{sec:intro}.

\subsection{ESA: Executable Sharing Agreement}

In existing systems, users are required to agree to long documents that are not understandable. Furthermore,
because these agreements are expressed in natural language, which can be ambiguous, it is not clear
what the effect of the agreements, or whether a certain company is truly in compliance. Instead,
we propose \textit{executable sharing agreement} (ESA), which have a well-defined unambiguous semantics; that is, whether a specific transaction complies with the agreement can be verified automatically. Furthermore, ESAs can automatically be converted into natural language unambiguously. This ensures that the contract is understandable to users and auditors.

Our ESA notation is inspired by the previously proposed ThingTalk language, designed originally for personal data sharing~\cite{commaimwut18}. The syntax of an ESA is as follows:
\begin{align*}
    \gamma, \pi(r, p)\texttt{:} \texttt{[}f_1, f_2, \ldots\texttt{]}~~\texttt{of}~~d, \pi(f_1, f_2, \ldots)
\end{align*}
This statement reads as follows: ``for consumer $\gamma$, the fields $f_1, f_2, \ldots$ from domain $d$ can be shared with
any \textit{requester} $r$ for purpose $p$, provided that the predicates $\pi(r, p)$ and $\pi(f_1, f_2, \ldots)$
are satisfied.

So, for example, to express that they are willing to share their abnormal PSA (a protein called prostate-specific antigen) with Stanford Medical Center, including their age and ethnicity but 
not their name, and only for research purposes, a patient named ``Bob'' might issue 
an ESA agreement of the form:

\begin{tabbing}
1234\=1234\=\kill
\>$\gamma = \text{``Bob''}, r = \text{``Stanford Medical Center''} \wedge p = \text{research}:$\\
\>\>$\left[\textit{age}, \textit{ethnicity}, \textit{PSA}\right]\texttt{ of }\text{EHR}, \textit{PSA} \ge 2$\\
\end{tabbing}

\subsubsection{Translating to and from Natural Language}

As consumers are not expected to understand formal languages, the ESA notation is designed to provide a natural language interface.  The sharing agreements can be translated from formal to natural language using a rule-based translation. The previous example can be expressed in natural language as follows:
\begin{quotation}
    ``Stanford Medical Center can read the age, ethnicity and PSA of Bob's EHR for research and if the PSA is greater than or equal to 2.''
\end{quotation}
While the automatically generated sentences can be verbose and clunky due to the rule-based translation, they are understandable, and they are guaranteed to correspond exactly to the code of the agreement.
Furthermore, the ESA notation is designed so that a user can define their own sharing agreement in natural language.  Previous work has shown that it is possible to automatically translate natural language access controls into attribute-based policies for personal data sharing~\cite{Genie, commaimwut18}, and the same semantic parsing technology is used here. Thorough analysis of semantic parsing is out of scope for this paper.

\subsection{ESA Enforcement}
\label{sec:esa-enforcement}

All writes to and reads from the database containing user data must go through the \Soteria interface which ensures compliance to all the sharing agreements, which represent user consent.   
Soteria automatically includes an \textit{owner} field for each row of the database.  Every database access is rewritten to include the sharing agreements constraints; it is timestamped and logged for later auditing.  

\subsubsection{Verification of SQL Queries for Auditing}

To prove compliance to an external auditor, \Soteria stores the requester, the purpose and the final query, right before it is issued to the database, including all the clauses related to the sharing agreements. Using the set of the sharing agreements in force at the time, the auditor can then formally verify that the query was compliant. Given a query from requester $r$ for purpose $p$ in the audit logs of the form:
$$\texttt{SELECT}~\bar{f}~\texttt{FROM}~t~\texttt{WHERE}~\pi$$
and sharing agreements of the form:
\begin{align*}
\gamma_1, \pi_1(r, p)\texttt{:} \left[f_1, f_2, \ldots\right]~~\texttt{of}~~t, \pi_1(f_1, f_2, \ldots)\\
\gamma_2, \pi_2(r, p)\texttt{:} \left[f_1, f_2, \ldots\right]~~\texttt{of}~~t, \pi_2(f_1, f_2, \ldots)\\
\ldots
\end{align*}
the query is compliant if and only if
\begin{align*}
\pi \models &\left(\gamma = \gamma_1 \wedge \pi_1(r, p) \wedge \pi_1(f_1, f_2, \ldots)\right) \vee \\
&\left(\gamma = \gamma_2 \wedge \pi_2(r, p) \wedge \pi_2(f_1, f_2, \ldots)\right) \vee \ldots
\end{align*}
This logical formula can be verified efficiently using a satisfiability modulo theory (SMT) solver~\cite{barrett2009satisfiability}.

\subsubsection{Formally Verified SQL Query}

To ensure that all queries are compliant, \Soteria uses the following algorithm to construct a query that is guaranteed by construction to satisfy the sharing agreements. Given a SQL query from requester $r$ for purpose $p$ of the form:
$$\texttt{SELECT}~\bar{f}~\texttt{FROM}~t~\texttt{WHERE}~\pi$$
and sharing agreements of the form:
\begin{align*}
\gamma_1, \pi_1(r, p)\texttt{:} \left[f_1, f_2, \ldots\right]~~\texttt{of}~~t, \pi_1(f_1, f_2, \ldots)\\
\gamma_2, \pi_2(r, p)\texttt{:} \left[f_1, f_2, \ldots\right]~~\texttt{of}~~t, \pi_2(f_1, f_2, \ldots)\\
\ldots
\end{align*}
\Soteria constructs a query of the form:
\begin{tabbing}
1234\=1234\=\kill
\>\texttt{SELECT}~$\bar{f} \cap \left\{f_1, f_2, \ldots\right\}$~\texttt{FROM}~$t$~\texttt{WHERE}~$\pi$~\texttt{AND}\\
\>\>$\left(\gamma = \gamma_1~\texttt{AND}~\pi_1(r, p)~\texttt{AND}~ \pi_1(f_1, f_2, \ldots)\right) \texttt{OR}$\\
\>\>$\left(\gamma = \gamma_2~\texttt{AND}~\pi_2(r, p)~\texttt{AND} \pi_2(f_1, f_2, \ldots)\right) \texttt{OR} \ldots$,
\end{tabbing}
where $\gamma$ is the field in the table containing the owner of that row.

It is possible to prove that the result set of this query is consistent with the sharing agreement. Only the fields allowed by the agreement are returned, and a row is included only if the predicates in the sharing agreement in force for the row owner are satisfied.

We note that the SQL query, while correct, might not be very efficient, as it includes at least one clause for each owner of the data in the database. In our experience, this is sufficient, as in practice the same sharing agreement is used by many different consumers, so clauses can be unified when the query is constructed. As consumers' preferences become more fine-grained, this might not be the case. Future work should investigate an optimization algorithm or an index structure suitable to queries of this form, so that both performance and formal correctness can be maintained.

\subsubsection{Enforcing Requester and Purpose}

A data requester can be an application user defined in CCPA or a company that stores user data.  The 
URM component of 
\Soteria does not verify the identity of the data requester. Such verification should be handled by the data transport protocol, and is out of scope of the \Soteria protocol. 
Figure~\ref{fig:soteria-arch} shows that applications outside the \Soteria
box handles user registration and authentication.  
\Soteria does not mandate any specific transport protocol; meaningful protocols could be REST, could be messaging-based~\cite{commaimwut18}, or could even be through physical media.  

Furthermore, \Soteria does not verify the purpose of the data access. The design assumes that at identification time, the data transport protocol will also compute the allowed purpose of data access. For example, using REST data transport, a medical data requester could be issued two different access tokens, one for clinical purposes and the other for research purposes. The correct purpose can be established through an existing business agreement between the data provider and data requester. If the data requester then uses the data in a manner inconsistent with the agreed purpose, the provider can use \Soteria to establish that the fault lies with the requester.  


\subsubsection{Enforcing Access on Write}

To support limiting what data is stored by a business entity such as AWS, queries that manipulate data (insert, update, delete) are also intercepted by \Soteria, and modified to match the ESA that limit the storage of data. Write queries that are not allowed by any ESA are not executed and are not logged.

To support permanent deletion of the data upon request by the consumer, the data itself (included in the \texttt{VALUES} and \texttt{SET} clauses of the insert and update queries) is masked in the audit log, as the audit log is immutable and append-only. For this reason, an ESA about data storage that depend on the specific values cannot be audited after the fact. \Soteria disallows such ESA: only the set of fields to write can be controlled. 

Continuing the previous example, the patient might issue a following the data storage ESA:
\begin{tabbing}
1234\=\kill
\>$\gamma = \text{``Bob''}: \left[\textit{age}, \textit{ethnicity}, \textit{PSA}, \textit{phone}, \textit{medication}\right]$\\
\>$\texttt{ of }\text{EHR.write}$\\
\end{tabbing}
This ESA allows age, ethnicity, PSA readings, contact information and current medications to be stored. It does not allow any other column to be stored, so the following query:
\begin{tabbing}
1234\=1234\=\kill
\texttt{INSERT INTO} EHR $\texttt{SET}~\textit{person} = \text{``Bob''},~\textit{age} = 52,$\\
\>\>$\textit{ethnicity}~= \text{white},~\textit{PSA}=1.5,~\textit{phone} = \text{``555-555-5555''},$\\ \>\>$\textit{smoker}=\texttt{TRUE},~\textit{consumesAlcohol}=\texttt{FALSE}, \ldots$\\
\end{tabbing}
would be rewritten as the following:
\begin{tabbing}
1234\=1234\=\kill
\texttt{INSERT INTO} EHR $\texttt{SET}~\textit{person} = \text{``Bob''},~\textit{age} = 52,$\\
\>\>$\textit{ethnicity}~= \text{white},~\textit{PSA}=1.5,~\textit{phone} = \text{``555-555-5555''},$\\
\>\>$\textit{smoker}=\texttt{NULL},~\textit{consumesAlcohol}=\texttt{NULL}, \ldots$\\
\end{tabbing}

Deletions are a special case of write, because they reduce the data that is stored about a specific customer rather than store new data. Hence, deletion is always allowed regardless of the current ESA. All deletions are logged, to prove that they were executed properly.

Additionally, upon revoking an ESA about data storage, some columns that were previously allowed might becomes disallowed. In that case, \Soteria overwrites the data to set those columns to $\texttt{NULL}$. In case all columns are now disallowed, such as when all ESAs for a customer are revoked, \Soteria deletes the row entirely. Both the database write (update or delete) and the ESA revocation are logged for later auditing.

\subsection{ESA Storage and Audit}
\label{sec:txns}

To support long-term auditability, as well as revocation of contracts, \Soteria makes use of a distributed ledger (blockchain) to track which sharing agreements are in force. The use of the blockchain provides a global ordering of all the events across all parties in the system; the events include issuing a sharing agreement, accessing the data, and revoking a sharing agreement. This global ordering ensures it is always possible to verify whether a sharing agreement was in force when a data transaction occurred, without disputes.

Sharing agreements can potentially be privacy-sensitive themselves; for example, the sharing agreement in the previous section can reveal that the patient is likely to be male. For this reason, \Soteria only stores the hash of the sharing agreement code, and the hash of each transaction, in the public blockchain. Upon request by a competent authority (e.g. under subpoena in a civil dispute), a data provider using \Soteria can reveal the full audit logs, including the full code of the sharing agreements, and the exact transactions performed. These logs can be then matched to the hashes stored in the public blockchain to verify that they were not tampered with.

\Soteria includes three types of events in the blockchain:
\begin{enumerate}
 
 
 \item {\em Sharing Agreement Deployment} (ESAD).
 An ESAD event occurs when a new sharing agreement is created between a consumer who controls the data, and a provider who owns it. An ESAD transaction on the blockchain (ESAD-TX) includes the PI owner's address, data provider' address, the hash of the agreement code, ESA deployment date, and ESA validity status (set to {\em true}).
 
 \item {\em Data Transaction} (DATA-TX).
 A data transaction indicates the transfer of data between a data provider and a data requester.
 DATA-TX records in the blockchain include the address of the data provider, the address of the data requester, and the hash of the exact query executed against the database. Note that, as described in the previous section, the exact query will include a reference to the consumers and their sharing constraints. Hence, given the raw audit logs, paired with the hash in the blockchain, it is possible to verify that the query was valid when executed.
 
 \item {\em Sharing Agreement Revocation} (ESAR).
 A sharing agreement can be revoked by a user, making it invalid. Since the ledger is append-only, a revocation is implemented by creating a new sharing agreement transaction with the validity status set to {\em false}. 
Note that a rectification request issued by a user is executed in two consecutive transactions,
an ESAR to revoke an existing consent and then an ESAD to create a new consent.
\end{enumerate}

\newpage
\section{DLT Consensus Protocols}
\label{sec:protocols}




\Soteria adopts blockchain technology to maintain its distributed ledgers 
among several institutions and users. 
To achieve high throughput and low latency, \Soteria uses a two-layer blockchain.
The base layer is decentralized similar to Ethereum, and its side chains 
are entrusted by a group of selected validators, known as {\em juries}. 
This two-layer approach does not compromise consistency between the data
maintained by \Soteria and the data on the providers' servers, as 
consistency can always be verified by users on the base-layer public blockchain.
Recall that the CAP theorem \cite{EB-cap-theorem} states that a distributed database 
system can only satisfy two of the three properties: 
consistency, availability, and partition tolerance.
While the base-layer ensures availability and partition tolerance,
the side chains ensure consistency and availability.
Using a permissioned side chain with a jury pool, 
\Soteria can improve both throughput and latency by limiting the 
number of juries.  
Note that permissioned blockchains are not the same as private blockchains. 
A permissioned blockchain gives the public the right to audit the system state but
only uses a limited number of juries to conduct validation, whereas
a private blockchain does not provide transparency to its system state.

\Soteria's ledger requires a consensus protocol 
since it supports contract revocation. 
A consensus protocol ensures that all validators agree on a unique order 
of transactions on the ledger.
Note that without this
strict access permission and access revocation order requirement, an append-only
log suffices to support the decentralized auditability requirement.

There are several types of consensus protocols,
each of which enjoys some pros and suffers some cons.
An application selects a particular consensus
protocol for its desired performance objectives (e.g., latency, throughput, and
consistency).  For instance, a protocol that guarantees immediate consistency may
trade latency and throughput for achieving the consistency objective.  
A protocol that requires 
just weak consistency or eventual consistency can achieve shorter latency and
higher throughput.  
Specifically, the PoX (proof-of-something) 
family protocols~\cite{bitcoin, King2012PPCoinPC, casper, nem} 
such as Proof-of-Work (PoW), Proof-of-Stake (PoS), and Proof-of-Importance (PoI)
offer timely consistency with good
network scalability but 
suffer from high latency and low transaction throughput. 
On the contrary, the BFT (Byzantine Fault Tolerance) family
protocols offer good performance with limited scalability w.r.t. 
the number of validators.
PoX is more suitable for permissionless blockchains (\Soteria's
base layer), where BFT for permissioned blockchains (\Soteria's side chains).

 In the remainder of this section, we survey representative BFT
consensus protocols including Tendermint~\cite{tendermintPaper}, Hashgraph~\cite{hashgraphWhitePaper}, HotStuff \sloppy~\cite{yin2018hotstuff}, and Stellar\cite{stellar}.  We summarize 
in Section~\ref{sec:qc} a qualitative comparison.

\subsection{Blockchain Overview and Terminologies}

\begin{figure}[t!]
\begin{center}
\includegraphics[width=8cm]{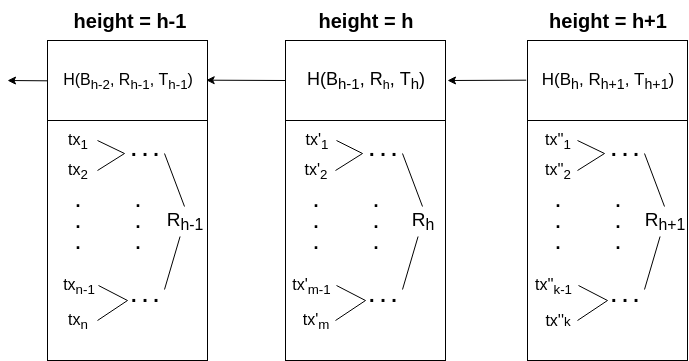}
\end{center}
\caption{Blocks with Hash Pointers}
\label{fig:Blocks}
\end{figure}

A blockchain of height $H$ is composed of a sequence of $H$ blocks.
Each block $h$, where $h = 0,\dots, H-1$, 
in a blockchain contains various numbers of 
transactions organized into a Merkle tree. The Merkle tree
root of the $h^{th}$ block, denoted as $R_{h}$, summarizes
the information in that block. Figure~\ref{fig:Blocks} shows 
an example blockchain.  The $h^{th}$ block, where 
$h \in [1,\dots,H-1]$ points to its 
previous block with pointer $B_{h}$. (The first 
block, or $h = 0$ is
the genesis block.)
$B_{h}$is the hash of three components: 
previous block hash $B_{h-1}$, the Merkle tree root of 
current block $R_{h}$, and some information from 
the $h^{th}$ block $T_{h}$ such as timestamp.

\newpage
\subsection{Base Chain Protocols}

Proof-of-something (PoX) protocols aim to support decentralized 
permissionless blockchains.  As mentioned in Section~\ref{sec:intro},
\Soteria uses a PoX blockchain as its base chain.  The choice
of a PoX protocol depends on the factors of cost and 
{\em inter-chain consistency latency}, 
which is defined as the time between when a transaction is committed on
a side-chain and the time when the root of the transaction's Merkle tree
is committed on the base chain. 
Inter-chain consistency is different from transaction consistency.
The former ensures a consistent public
view of transactions for the purpose of decentralized audits, whereas
the latter ensures the validity of individual transactions.
Transactions committed on a side chain guarantees local 
order on that chain to be consistent. The main chain, does not
guarantee total order all transactions across all side chains. 
Once \Soteria enforces that a contract revocation must
take place on the same side chain where the contract was agreed upon
and validated, side chain consistency guarantee suffices.

For an access revocation transaction, a side chain of \Soteria
can ensure transaction-commit latency to be within seconds. 
In other words, once a user revokes a prior permission to access her data, 
her data will be inaccessible within seconds. For the purpose of auditing,
latency is defined as the time required to update side-chain information
on the main chain.  This inter-chain consistency latency is 
the time between when a transaction commits 
on a side chain and when the root of the transaction's Merkle tree
is hashed onto the main chain.  Although \Soteria 
can only guarantee inter-chain consistency  
within minutes, this suffices for the auditing purpose required 
by regulations\footnote{E.g., the CCPA 
announced in January 
2020 requires a data holder to respond
to an audit request in $45$ days.}.
A PoS (proof-of-stake) protocol such as Ethereum satisfies
latency in minutes at a relatively low cost (compared to e.g., Bitcoin).
Therefore, \Soteria uses Ethereum as its main chain.

\subsection{Side Chain Protocols}
\label{sec:side-protocols}

Byzantine Fault Tolerance (BFT) protocols are
more efficiently but relatively small in scale in terms
of the size of a jury pool (or the number
of voting members) than public blockchains. 
One major milestone of BFT is the deployment of Practical  
Byzantine Fault Tolerance (PBFT), which is the first 
implementation that works correctly in an asynchronous 
system such as the Internet~\cite{pbft}. 

PBFT has two modes, normal mode and view-change mode. 
In normal mode, a leader proposes a candidate value 
to the other replicas in the pre-prepare phase. 
PBFT then goes through two successive
voting phrases: prepare and commit. 

If the candidate value is accepted by a replica $p_{i}$, as known as a validator, 
$p_{i}$ then enters the prepare phase and broadcasts a prepare message to others 
consisting the candidate value.
Once $p_{i}$ collects enough messages, i.e., $2f+1$ messages 
over $3f+1$ replicas ($f$ denotes the number of Byzantine nodes) and 
agree on the same value, it enters into commit phase.
In the commit phase, replicas conduct an election similar to the one in 
the prepare phase to agree that more than $2f$ replicas will 
write the candidate value into their respective databases.

To prevent indefinitely waiting, 
$p_{i}$ transitions to view-change mode if a timeout is triggered. In view-change mode, 
replicas start a new view to elect a new leader
by sending view-change messages.

\subsection{BFT/PBFT Protocols and Comparison}

Since BFT/PBFT family protocols meet CA properties 
and performance requirements of \Soteria's side chains,
we survey and compare four representative protocols belonging to this family.
For qualitative comparison documented in Section~\ref{sec:qc},  
we survey and present four well-documented protocols, {\em Tendermint}, 
{\em Stellar}, {\em HotStuff}, 
and {\em Hashgraph}.  
Enhancements proposed by some other protocols (e.g.,
\cite{bftsmart, algorand, honeyBadger, sbft}) stem from these four protocols.
In Section~\ref{sec:experiments} we benchmark two deployed and
battle-tested\footnote{It is essential for a protocol to have been battle-tested 
in real-world deployments to ensure it reaches its claimed performance and
fault tolerance.} protocols, {\em Tendermint} and {\em Stellar}, to 
compare throughput and latency quantitatively.  
For each protocol, we detail its algorithms and discuss 
its advantages and shortcomings.

 \subsubsection{Tendermint}

 Tendermint is a leader-based BFT approach that does not conduct leader selection, aiming to
 reduce the number of communication IOs. 
 Unlike PBFT which is composed of two modes, normal mode and view-change mode, 
 Tendermint can frequently change leaders (proposers) and make consensus progress 
 under a single mode by using a predefined leader selection function. 
 In a nutshell, the idea of Tendermint to embed view-change mode into normal mode in PBFT.
 With this idea, Tendermint reduces the communication complexity by an order of magnitude compared 
 to PBFT: from $\mathcal{O}\left(N^3\right)$ in PBFT to $\mathcal{O}\left(N^2\right)$, where $N$ denotes
 the number of validators.


For a new block, validators go through several {\em rounds} utill they reach a consensus on
that block. Each round consists of four phases, 
{\em propose}, {\em prevote}, {\em precommit}, and {\em commit}. 
Tendermint's voting process is the same as PBFT's, in which each validator collects votes from 
the others between {\em propose} and {\em prevote} phases and between {\em precommit} and {\em commit}
phases. Let $f$ denote the number of Byzantine fault nodes.
When over $N-f$ of the same vote have been collected by a validator, the validator 
switches from the first phase to the second phase.
Furthermore, Tendermint implements the following three functions to
ensure safety and liveness properties.

\begin{enumerate}
\item {\em Predefined leader selection function}. At the beginning of a new round (propose phase), validators choose a new leader with this function, 
which is weighted round-robin against these validators' voting power.

\item {\em Timeout function}. 

As a partially synchronous model, Tendermint triggers a timeout in between every two phases if a validator cannot receive enough votes within a specified duration. This timeout prevents a validator from waiting forever. To adapt to network delay, Tendermint uses a timeout function \(timeout\left(r\right) = timeout\left(r-1\right) + r * \Delta \), where $r$ is the round number and $\Delta$ is a 
configurable constant.  The initial timeout value ($r = 0$) is set by  
a configurable constant.  When a timeout occurs, a new round starts 
with a longer timeout value.



\item {\em Voting lock mechanism}. 
To validate a block, each validator maintains two values, 
a {\em locked value} and a {\em valid value}. 
If a validator $p_{i}$ sends a {\em precommit} message with a valid $v$ value ($v \ne nil$), it locks $v$ as its locked value. 
If a validator $p_{i}$ receives $N-f$ {\em prevote} messages with the same valid value $v$, it updates its valid value with $v$. Also, if the proposer already has had a valid value, it proposes only this value to other validators. That valid value is regarded as the most recent possible value.
Once a locked value $v_{locked, i}$ is updated by $p_{i}$ in round $r_{l}$, it can only vote for $v_{locked, i}$ in the following rounds unless validator $p_{i}$ receives a valid value $v_{valid, j}$ from the new proposer $p_{j}$ and satisfies $r_{k} > r_{l}$, where $r_{k}$ is the round where $p_{j}$ updated its valid value to $v_{valid, j}$. 
Validator $p_{i}$ can unlock $v_{locked, i}$ and vote for $v_{valid, j}$ in the {\em prevote} phase.
\end{enumerate}

 The intuition of Tendermint is that at block $h$ at least one correct validator $p_{i}$ proposed $v_{valid, i}$ in round $r{l}$ is acceptable by $N-f-1$ correct validators. In addition, given any of these $N-f-1$ correct validators $p_{j}$ and its locked value $v_{locked, j}$ in round $r_{k}$, one of the following two conditions is always satisfied, $v_{valid, i} = v_{locked, j}$ or $r_{l} > r_{k}$.
 This means that every correct validator can always send {\em precommit} message with $v_{valid, i}$ in the same round and commit a block on block $h$. Therefore, correct validators can always commit the same valid value, and hence guarantees the safety property \cite{tendermintPaper}. 
 Besides, contrary to PBFT, adapting these mechanisms does not require forwarding 
 additional information (e.g., sending $N-f$ valid check point messages) to the new proposer.
 
 Tendermint also can be improved by using threshold signatures \cite{yin2018hotstuff} for reducing message complexity to improve the scalability. In addition, the protocol has been proven to achieve fairness under certain conditions \cite{correctnessFairnessTendermint}. However, since each change of leader requires waiting for a known upper bound of the network delay, which may exceed the maximum actual network delay, Tendermint does not enjoy {\em optimistic responsiveness} defined in \cite{thunderella}.

The pros and cons of Tendermint are:
\begin{itemize}
    \item Pros: With a small jury of trusted validators, Tendermint achieves low latency with safety.
    \item Cons: Its number of communications is larger than the other protocols, and therefore, the size of a jury cannot be too large.  This makes Tendermint a good protocol for 
    DataXchange's side chains, but
    not desirable to be its main chain.
\end{itemize}
\subsubsection{Stellar}

The Stellar consensus protocol (SCP) is composed of two protocols, the {\em nomination} protocol and {\em ballot} protocol. These protocols reach a consensus by using federated voting, which supports flexible trust. With flexible trust, users can choose any parties they trust to join the consensus process without a central authority. 

More formally, every validator can select a set of validators to establish
a quorum slice. For example, validator $p_{i}$ selects a set of validators denoted as $L$ to form 
quorum slice $QS_{\left(i,L\right)}$. Validator $p_{i}$ can create multiple 
quorum slices, or $S_{i} =  \{Q_{\left(i,L_1\right)}, QS_{\left(i,L_2\right)}, \dots ,QS_{\left(i,L_n\right)}\}$. 
A quorum slice is defined as a set of validators sufficient to reach an agreement.


SCP assumes {\em quorum intersection} in that two quorum slices share at least one well-behaved node.
SCP is a leaderless PBFT protocol, and it does not have the {\em pre-prepare} phase.
However, SCP adds an {\em accept} phase to PPBFT to design 
its {\em federated voting} scheme to reach 
global consensus. Thus, federated voting includes three phases following the order 
of {\em prepare}, {\em accept}, and {\em confirm}. 

\begin{enumerate}
\item {\em Prepare}. Once a new instance $v$ has been created, 
$p_{i}$ sends a message with value $v$ to whomever trusts $p_{i}$.
Once $p_{i}$ votes for $v$, it will not vote for other values.

\item {\em Accept}. Validator $p_{i}$ receives messages from the validators in its quorum slices $QS_{i}$. Validator $p_{i}$ can change its voted value in the {\em prepare} phase if the majority of $p_{i}$'s {\em blocking set} vote for another value $v^{\prime}$ in the {\em prepare} phase. 
Otherwise, $p_{i}$ still stands for the value $v$ it voted in the {\em accept} phase. 
The {\em blocking set} of $p_{i}$, $B_{i}$, is defined as:
for each quorum slice of $p_{i}$, $QS_{\left(i,n\right)} \in QS_{i}$, 
such that \(\forall QS_{\left(i,n\right)} \cap B_{i} \neq \emptyset\), 
which means $B_{i}$ contains at least one node in every quorum slice of $p_{i}$.

\item {\em Confirm}. $p_{i}$ broadcasts the value that it voted in the {\em accept} phase, 
thereby committing to that value.
\end{enumerate}

The reason for the {\em accept} phase
is that if the value voted by a validator in the {\em prepare} phase 
is different from the value voted by the majority of its neighbors, the neighbors can convince the validator to accept the other value. Therefore, although the SCP network structure can be configured the
same as that of PBFT, SCP does not request a validator 
to commit a block with $N-f$ of the same votes in any phases. Instead,
validator $p_{i}$ only conveys messages to those who list $p_{i}$ in their quorum slices. 



In summary, SCP provides users with the flexibility to make their own quorums, and its use of 
federated voting can lead to consensus and enjoy the safety property based on the assumptions listed in the white paper \cite{stellar}. These assumptions may limit the freedom of network extension, since validators require adequate settings of quorum slices. 

The pros and cons of SCP are:
\begin{itemize}
    \item Pros: First, users have the privilege of selecting quorum slices composed of parties 
    that they can trust. Second, based on the BFT-based protocol, SCP achieves low latency with safety.
    \item Cons: Setting up quorum slices may introduce network structure errors.
\end{itemize}
\subsubsection{HotStuff}
\label{subsubsec:hotStuff}

HotStuff is a leader-based BFT protocol in partial synchronous 
settings for reaching consensus. 
HotStuff is a four-phase consensus protocol including {\em prepare}, {\em pre-commit}, 
{\em commit}, and {\em decide}.  (For details of
these phases please refer to \cite{yin2018hotstuff}.) 
HotSuff fulfills two properties: 
linearity and optimistic responsiveness:

\begin{itemize}
	\item {\em Linearity}.  The consensus algorithm has a linear communication complexity for committing a block in one round. Two scenarios must be considered.
	First, if a designated proposer (leader) is correct, the proposed block only takes  $\mathcal{O}\left(N\right)$ messages ($N$ was defined as the number of
	validators) to be committed by the majority. Otherwise, a new proposer must be elected, known as view-change as defined in Section~\ref{sec:side-protocols}, in 
	the limit of $\mathcal{O}\left(N\right)$ messages as well.  
	In the worst case, the total communication cost is 
	$\mathcal{O}\left(N^2\right)$ for at least $f$ ($f \propto N$) times proposer selection.
	We will shortly explain how 
	this linearity is achieved by HotStuff.
	
	\item{\em Optimistic responsiveness}. Any correct proposer $p_{i}$ only has to wait for a delay of $N-f$ messages to ensure that the candidate value $v$ that $p_{i}$ proposed can be committed (in view change mode).
	This message delay is independent of the known set upper bound $\delta$ on the network's delay \cite{thunderella}. 
\end{itemize}

HotStuff uses two techniques to achieve the {\em linearity} property. First, it adopts a
simpler message transmission pattern.
Instead of letting each validator broadcast messages to the others in any consensus phases, 
HotStuff makes all validators send signed messages to a designated leader. Then the leader 
integrates messages into an agreement. If $N-f$ validator agree on
the same content, the result is broadcast to the other validators to be voted on in the next phase. 


Second, HotStuff employs the {\em threshold signature} scheme to reduce the message complexity.
To verify whether an agreement broadcasted by the leader is actually signed by validators, the message should contain $N-f$ valid messages signed by validators.
These $N-f$ attached signatures incur $\mathcal{O}\left(N^2\right)$ communication complexity in the consensus process.
However, with the threshold signature technique, an agreement which only carries one signature can be verified by validators.
With these two schemes, HotStuff can satisfy linearity.

In the threshold signature scheme, all validators hold a common public key $PUB$, but each validator
denoted as $p_{i}$ owns a distinct private key $PRI_{i}$.
A threshold signature $\rho$ can be defined as:
\(\rho \leftarrow combine\left(m, S\right)\),
where \(S = \{\rho_i \mid p_{i} \in V\}\), V is the set of validators where $|V| = N$,
and \(|S| \geq N-f \wedge \rho_{i} \leftarrow sign\left(m, PRI_{i}\right) \) where $sign\left(m, PRI_{i}\right)$ means validator $p_{i}$ signs the message $m$ with its private key $PRI_{i}$ to create the partial signature $p_{i}$.

Two trade-offs are made in order to achieve linearity. First, compared to 
three-phase protocols, HotStuff adds an additional phase to each view, which causes a small amount of latency~\cite{yin2018hotstuff}. 
Second, the threshold signature can cost more computational resources. Taking Rivest–Shamir–Adleman (RSA) based implementations of threshold signature schemes as an example, given a $\left(t, n\right)$ threshold signature scheme, at lease $t$ or more participants in a group of
size is $n$ can generate a valid signature collaboratively. 
According to the work of ~\cite{thresholdSigSurvey}, the computation 
complexity can range from $t \times T_{RSA}$ to $3t \times T_{RSA}$ for an individual 
signature generation and verification, depending on the feature design choices. ($T_{RSA}$ represents 
the time for computing exponent in an RSA-type scheme.) 
Also, in Elliptic curve based implementations, generating an individual signature still requires $t$ times of $T_{EC}$, the time for computing a Elliptic curve~\cite{ChangThresholdSig}.
In HotStuff, $\left(t,n\right)$ is $\left(f,N\right)$, which means that the threshold signature scheme could cost $f$ times of RSA-type schemes or Elliptic-curve-type schemes.

The pros and cons of HotStuff are:
\begin{itemize}
    \item Pros: HotStuff optimizes the message complexity. 
    Chained HotStuff also simplifies the algorithm and allows for more frequent leader rotation.
    \item Cons: HotStuff adopts a more sophisticated cryptography mechanism, which takes longer time
    to reach a consensus, compared to RSA-type schemes or Elliptic-curve-type schemes.
\end{itemize}

\begin{table*}[htbp]
\centering
\begin{tabular}{|c||c|c|c|c|}
\hline
                                         & \textbf{Tendermint}        & \textbf{Stellar}       & \textbf{HotStuff}                                        & \textbf{Hashgraph}\\ \hline
\textbf{Timing Model}                    & Partial Synchronous        & Partial Synchronous    & Partial Synchronous                                      & Asynchronous\\\hline
\textbf{Key Design Goals}                & Single mode mechanism      & Flexible trust         & \makecell{1. Linearity \\2. Optimistic responsiveness}   & High throughput\\\hline
\textbf{Fault Tolerance}           & $\leq \frac{2}{3}$ Voting Power  & -                      & $\leq \frac{2}{3}$ Voting Power             & $\leq \frac{2}{3}$ Voting Power\\\hline
\textbf{Message Complexity}             & $\mathcal{O}\left(N^2\right)$ - $\mathcal{O}\left(N^3\right)$ & -  & $\mathcal{O}\left(N\right)$ - $\mathcal{O}\left(N^2\right)$ & $\mathcal{O}\left(N^2\right)$\\\hline
\textbf{Scalability}                     & $100$-$1,000$                   & -                      &$\geq 100$                                                 & $\geq 1,000$\\\hline
\textbf{Validator Bound}                 & $64$                       & $43$                   & $128$                                                     & $128$\\\hline
\textbf{Throughput (tx/s)}               & $4k$                       & -                      & $10k$                                                     & $\geq 50k$ \\\hline
\textbf{Latency (s)}                     & $5$                        & $1.3$                   & $10$                                                     & $\geq 10$\\\hline
\textbf{Benchmark Setup}                 & AWS t2.medium              & AWS c5d.9xlarge        & AWS c5.4xlarge                                            & AWS m4.4xlarge\\\hline
\end{tabular}
\caption{Consensus comparative analysis \protect\cite{tendermintThesis, stellar_2, yin2018hotstuff, hashgraphWhitePaper}.}
\label{tab:comparison}
\end{table*}

\subsubsection{Hashgraph}

The Hashgraph consensus protocol (HCP) is a BFT solution in a completely asynchronous model. 
HCP consists of a hashgraph data structure, a {\em gossip about gossip} data-transfer 
algorithm, and a {\em virtual voting} mechanism.  We describe these components as follows.

\begin{itemize}
\item {\em Hashgraph}. A hashgraph is a directed acyclic graph (DAG). Each validator maintains a local copy
of the hashgraph. A graph node is known as an {\em event},
which records a timestamp, transactions, and two pointers, 
one pointing to {\em self-parent} event and the other to {\em other-parent} event. 
Creating a new event can be regarded as proposing a new candidate value.

\item {\em Gossip about gossip}. 
Gossip about gossip is a process of synchronizing two validators' local hashgraph. 
Validator $p_{i}$ randomly selects a peer $p_{j}$ to transfer events that $p_{j}$ does 
not known yet. Validator $p_{j}$ only accepts events with valid signatures and places those events 
on its hashgraph. Finally, $p_{j}$ creates a new event $e$ whose pointers link to 
both $p_{i}$ and $p_{j}$. 
Event $e$ also indicates the fact that $p_{i}$ has shaken hands with $p_{j}$.

\item {\em Virtual voting}. 


The virtual voting process is similar to the process of PBFT in the normal mode.
If validator $p_{i}$ verifies the transactions wrapped in event $e_{t-1}$ to be valid, 
validator $p_{i}$ creates a new event $e_{t}$ and points $e_{t}$ to $e_{t-1}$. 
Event $e_{t}$ will be transferred when the next round of gossip about gossip process starts. 
Otherwise, $p_{i}$ drops $e_{t-1}$. 
From validator $p_{j}$'s view, if $e_{t}$ has been 
sent to $N-f$ validators, HCP defines that $p_{j}$ can strongly 
see $e_{t-1}$. The concept of `strongly seeing' is similar to `going to the next phase' in PBFT.
Moreover, there are two voting rounds, just like prepare and commit, to confirm events.
Therefore, every validator has to send each voting message to other validators 
directly or indirectly, which means the message complexity is still $\mathcal{O}\left(N^2\right)$


\item {\em Coin flip}. According to the FLP impossibility, it is
impossible for a consensus algorithm running on an asynchronous network to achieve both safety and liveness under node failure~\cite{Impossibility}. Therefore, to avoid the impossibility in an asynchronous model, Hashgraph consensus protocol adopts randomized algorithm by allowing validators to flip coins. That is, a validator periodically votes pseudorandomly depending on the middle bit of signatures of events. Furthermore, validators do not broadcast coin-flip results since they can tally votes based on their local copies.
\end{itemize}

In HCP's white paper \cite{hashgraphWhitePaper}, the safety property is proofed. 
For liveness, HCP guarantees only termination of all non-faulty processes with probability one 
in its asynchronous model. 
However, there is no upper bound of time to reach a consensus.
In addition, a local coin-flip protocol such as Ben-Or \cite{benor} requires 
exponential expected time to converge in the worst case \cite{Aspnes2003}.
Therefore, if Byzantine nodes manipulate the gossip protocol and it takes validators 
numerous rounds to reach consensus, the transaction latency can suffer a great deal.


The pros and cons of Hashgraph are:
\begin{itemize}
    \item Pros: The message complexity is reduced by an order compared to PBFT. With a small-scale network, HCP can achieve low latency.
    \item Cons: Latency can be long in a large-scale network.
\end{itemize}

\subsection{Qualitative Evaluation}
\label{sec:qc}

In order to provide a qualitative comparison between these protocols we 
have enumerated, we collected data from relevant publications~\cite{tendermintThesis, yin2018hotstuff, hashgraphWhitePaper, stellar_2}. 
Table \ref{tab:comparison} presents a summary including
the following properties:
\begin{itemize}
\item Timing model: timing assumptions of different models including 
synchronous, asynchronous, and partial synchronous.

\item Design goals: the primary performance goals that a consensus algorithm 
was designed to achieve. This provides the contextual information for
evaluating a protocol.  (A protocol designed for achieving low latency should not be
derided for its weaknesses in other performance metrics.)

\item Fault tolerance: the upper bound of 
faulty nodes (or weighted votes) that can cause system failure.

\item Message complexity: the overall message complexity to 
commit a block. $N$ denotes the number of validators.

\item Scalability: the number of validators that the consensus protocol claims 
to be able to participate in the consensus process. 

\item Validator bound: the largest number of validators 
that can participate in a consensus protocol.

\item Throughput: the number of transactions per second that can be committed by the majority of validators under the largest number of validators.

\item Latency: the average delay before a transaction being committed under the largest number of validators.
\end{itemize}



Table~\ref{tab:comparison} shows only the performance data under the largest 
number of validators that a protocol can support given its own hardware
configuration and assumptions. 
In Section~\ref{sec:experiments}, we presents our own experiments
attempting to do a relatively fair comparison.

Note that Stellar is absent in most of the fields 
in Table~\ref{tab:comparison} for two reasons. 
First, fault tolerance and message complexity are highly correlated with the 
configuration of each validator in the Stellar network. 
This flexible configuration makes it difficult to analyze Stellar 
without knowing its network structure. Second, the work of \cite{stellar_2} focuses
only on reducing latency and assumes that throughput can be improved by adding hardware. 
Also note that thought Table~\ref{tab:comparison} does not provide
an apple-to-apple comparison, it
provides a birds-eye view on representative protocols' 
characteristics.
Although all protocols do their best to optimize performance, we can still
observe a general tradeoff between latency and throughput. For instance,
HCP attempts to optimize throughput, but its latency is higher than
the others.  Conversely, Tendermint enjoys low latency but suffers 
from relatively low throughput.

\newpage
\section{Experiments}
\label{sec:experiments}

\begin{figure*}[ht]%
   \centering
    \subfloat[Throughput vs. Input Rate.]
    {{\includegraphics[width=8.3cm]{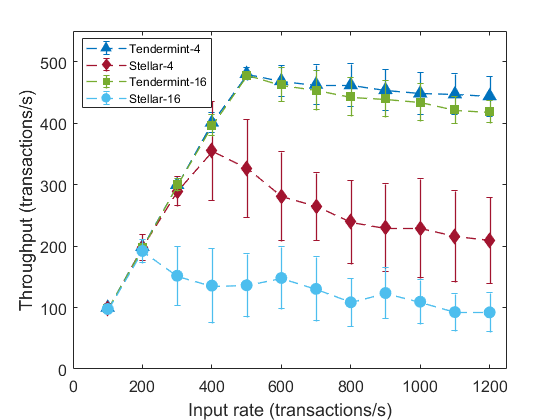} }}%
    \qquad
    \subfloat[Latency vs. Input Rate.] {{\includegraphics[width=8.3cm]{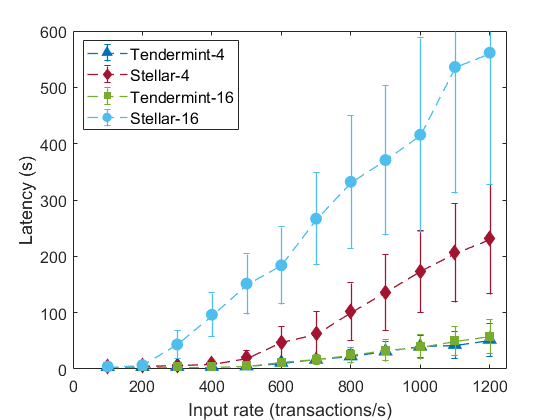} }}%
    \caption{Performance of Tendermint and Stellar with $4$ and $16$ Nodes (validators).}%
    \label{fig:performanceComparison}%
\end{figure*}

Our experiments were designed to evaluate tradeoffs between latency and throughput under
different hardware and software configurations. Unfortunately, since 
there is no stable implementation of Hashgraph or Hotstuff\footnote{Hashgraph does
not provide open-source release. Hotstuff's open-source code is not free of bugs as of Q4 2019 to put into production.} to date,
we evaluated only
the latest versions of Tendermint and Stellar. 
We adopted Tendermint v0.32.1 and Stellar v12.0.0 and deployed
them on Google Cloud Platform (GCP). 
For Tendermint and Stellar, respectively, we created up to $64$ validators, 
and assigned an 8-core Intel(R) Xeon(R) $2.20$ GHz CPU with $7.2$ GB memory to each validator.
Validators were configured into a fully connected topology.
To measure realistic network latency, we deployed validators on Google 
servers in different geographical locations (Taiwan, Singapore, 
Belgium, and Columbia). Validators were distributed equally to these locations. 
For example, in a $64$-validator implementation, $16$ validators were allocated 
to each of the four geographic locations.
 
\subsection{Metrics}

We define throughput and latency based on 
an end-user's perspective.
Given a set of $N$ validators $V=\{p_{1},p_{2},\dots,p_{N}\}$, 
throughput denoted as $\textrm{TPS}_{\textrm{avg}}$ is written as
\begin{equation}
    \textrm{TPS}_{\textrm{avg}} = \frac{\Sigma_{i=1}^{N} \textrm{TPS}_{\textrm{avg}}^{i}}{N},
\end{equation}
where $\textrm{TPS}_{\textrm{avg}}^{i}$ denotes the average throughput of 
validator $p_{i}$, who owns a sequence of blocks $B^i=\{b_0^i, b_1^i, \dots, b_{H^i}\}$.
In turn, $\textrm{TPS}_{h}^{i}$ 
(throughput of the 
${h+1}^{th}$ block of validator $p_{i}$) is the average
throughput of all transactions in that block, which is obtained by the number of transactions in
the block divided by the commit interval between $b_{h-1}^i$ and $b_{h}^i$, which 
is denoted as $\Delta t_{h}^i$, or
\begin{align*}
    \textrm{TPS}_{\textrm{avg}}^{i} = \frac{\Sigma_{h=0}^{H_i} 
    \textrm{TPS}_{h}^{i}}{H_{i}}, {\enspace \textrm{where} \enspace}
    \textrm{TPS}_{h}^{i} = \frac{|T_{h}^i|}{\Delta t_{h}^i}.
\end{align*}

The latency of a transaction is the time between when the transaction
is submitted by a client and when the transaction is committed in a block.
Each transaction $j$ may have different latency in $V$ validators' 
databases, $\textrm{Latency}_{i}^{j}$, $\forall p_{i} \in V$. We 
thus denote $\textrm{Latency}_{\textrm{med}}^{j}$ as the median 
latency of transaction $j$ among all $V$ validators.
The average latency of $T$ transactions denoted as $\textrm{Latency}_{\textrm{avg}}$ can then be written as
\begin{equation}
    \textrm{Latency}_{\textrm{avg}} = \frac{\Sigma_{j=1}^{T} \textrm{Latency}_{\textrm{med}}^{j}}{T}.
\end{equation}

\subsection{Parameter Settings}

Several parameters affect the block commit time, including 
maximum block size, transaction size,  
commit interval, transaction weight, 
and quorum-slice size~\cite{yin2018hotstuff, tendermintThesis}. 

\begin{itemize}
    \item {\em Maximum block size}. The maximum number of transactions that
    can be included in a block as the block is confirmed.
    \item {\em Transaction size}. The capacity of a transaction in bytes.
    \item {\em Minimal commit interval}. The minimum time increment between consecutive blocks. 
    A minimum commit interval may prolong block commit time if an implementation intends to produce a block whose latency is smaller than the commit interval. However, we assume that a block's latency, which will be no longer than the commit interval, is an acceptable delay.
    \item {\em Transaction weight}. The probability of a transaction being committed in a block. All transactions have the same transaction weight in our experiments.
    \item {\em Quorum slice}. In Stellar, each validator has its quorum slice \cite{stellar}. 
    To achieve a fully connected network, each validator's quorum slice is comprised of
    the rest of the validators so that all validators join the quorum and there is no dispensable set.
\end{itemize}
\begin{table}[h]
\centering
\small
\begin{tabular}{l||c|c|c}
 & \textbf{Maximum block size} & \textbf{Transaction size} & \textbf{Commit interval} \\ \hline
 & $3,000$ transactions/ block     & $212$ bytes                 & $5$ seconds
\end{tabular}
\caption{Parameter Settings for Tendermint and Stellar}
\label{tab:parameter_setting}
\end{table}

We used the parameter settings listed in Table~\ref{tab:parameter_setting}
to compare Tendermint and Stellar,
Both implementations were issued the same workload of 
a large number of synthetic transactions.
To simulate realistic client behavior, 
we used a separate node to submit transactions 
to validators continuously for $60$ seconds 
\footnote{Our initial experiments indicated that the block commit 
intervals of both Tendermint and Stellar can be stabilized in $60$ seconds.
To conserve AWS budget, we therefore ran all experiments for $60$ seconds.}.
This node sent transactions in different input rates up to $1,200$ 
transactions/s in a round-robin fashion.

\subsection{Throughput and Latency Comparison}

Figure~\ref{fig:performanceComparison} presents the performance of both implementations
with $4$ and $16$ validators (or CPU nodes).  (Later in the scalability study, we
present performance with larger number of nodes.)
With $4$ validators (or nodes), Tendermint and Stellar have 
nearly the same performance when the input rate is under $300$. However, when input rates exceed $400$ transactions/s, Tendermint achieves higher throughput and lower latency. The gap between 
the two protocols widens as the input rate is further increased. 

\begin{figure}[h!]
    \centering
    \includegraphics[width=8.8cm]{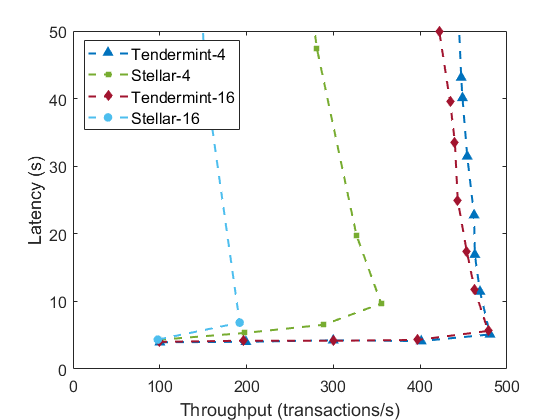}
    \caption{Throughput vs. latency of Tendermint and Stellar with $4$ and $16$ nodes. In each dashed line,
    the consecutive vertices represent the input rate increases,
    starting from 100 transactions/s and increasing by 100 transactions/s each time.}
    \label{fig:tps_latency}
\end{figure}

Furthermore, we examine peak performance, which is defined as 
the highest throughput with the lowest latency.
In Figure~\ref{fig:tps_latency}, we plot throughput ($x$-axis)
and latency ($y$-axis) tradeoff for $4$ and $16$ validators of
Tendermint and Stellar, respectively.  The nodes on each line represent
the input rates from $100$ per second onward by increments of $100$ per second.
As an example, for Stellar-$4$ (Stellar with $4$ validators), its throughput initially steadily increases from an input rate of $100$ to $400$ transactions/s, but then begins to degrade as the input rate increases beyond $400$ transactions/s.

Meanwhile, latency grows rapidly (beyond the
upper bound of the figure after input rate is larger than $600$ transactions/s) when
throughput starts to degrade due to saturation of 
system resources. This resource-saturation problem suggests to us that 
admission control and load balancing
must be implemented in a production system.

The peak performance of Stellar-$4$ is around 350 transactions/s with 10 seconds latency at an input rate of $400$ transactions/s. Tendermint-4 can reach a performance of $490$ transactions/s with $5$ seconds latency at an input rate of $500$ transactions/s.
Increasing the number of validators from $4$
to $16$, the performance gap widens.

In summary, in low traffic situations (under $300$ transactions/s), 
Tendermint-$4$ and Stellar-$4$ perform at about the same level. 
However, with heavier traffic, Tendermint outperforms Stellar.

\subsection{Scalability}
\label{subsec:scalability}

\begin{figure}[t!]%
    \centering
    \subfloat[Throughput vs. Input Rate.] {{\includegraphics[width=8.8cm]{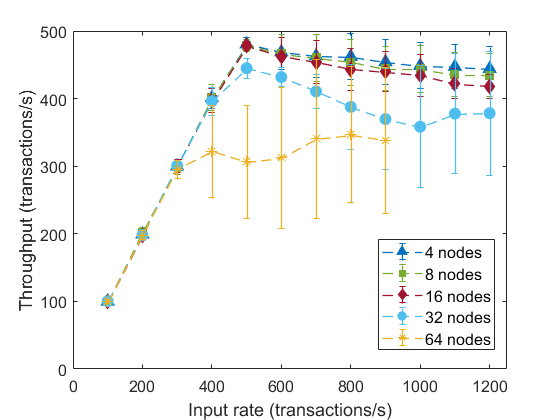} }}%
    \qquad \\
    \subfloat[Latency vs. Input Rate.] {{\includegraphics[width=8.8cm]{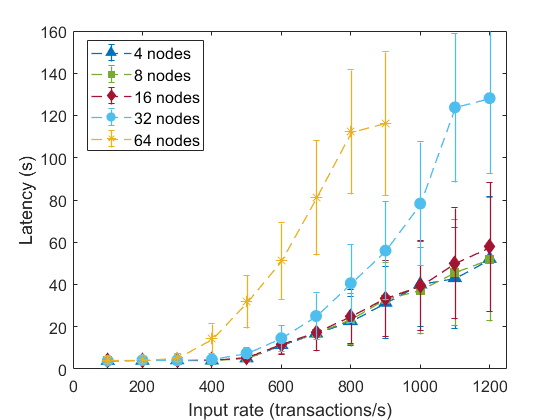} }}%
    \caption{Tendermint Performance with $4$ to $64$ Nodes (Validators)}%
    \label{fig:TendermintExperimentData}%
\end{figure}

\begin{figure*}[h]%
    \centering
    \subfloat[Throughput vs. $\#$ of Nodes.] {{\includegraphics[width=8.3cm]{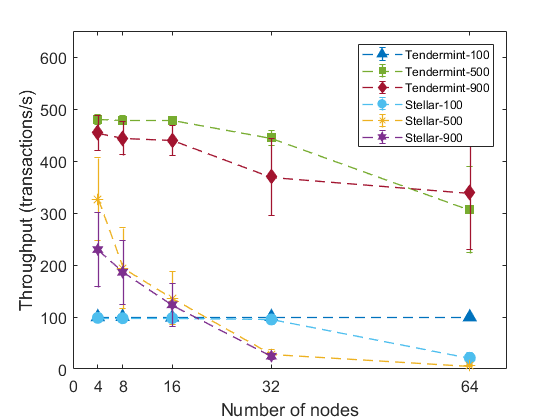} }}%
    \qquad
    \subfloat[Latency vs. $\#$ of Nodes.] {{\includegraphics[width=8.3cm]{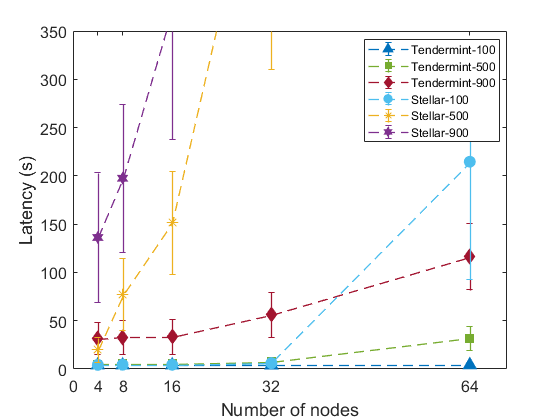} }}%
    \caption{Performance of Tendermint vs. Stellar at Input Rates $100$, $500$, and $900$.}%
    \label{fig:scalabilityComparison}%
\end{figure*}

This section presents results of our
scalability study.  Figure~\ref{fig:TendermintExperimentData} 
depicts Tendermint's through and latency
at input rates increased from $200$ to $1,200$ transactions/s.
Stellar's performance exhibits the same patents as Tendermint's, and
we do not separately present its data.
We report Tendermint vs. Stellar comparison in 
Figure~\ref{fig:scalabilityComparison}.

From Figure~\ref{fig:TendermintExperimentData}[a], we observe that 
the throughput of Tendermint eventually saturates and degrades after the input rate has
reached the capacity of the system.  Throughput degradation occurs at a lower
input rate when the number of validators (the number of nodes) are larger.
The degradation point of $16$-node Tendermint is at input rate $450$ transactions/s,
while the degradation point of $64$-node configuration starts at $250$.
Figure~\ref{fig:TendermintExperimentData}[b] depicts latency versus input rate.
At the same input rate when throughput starts to degrade, the latency of Tendermint
also increases drastically. For smaller configurations from $4$ to $16$ nodes
latency reaches $50$ seconds at input rate $1,200$, whereas for lager configurations $32$ and $64$
latency reaches two minutes. 

It is expected that system capacity eventually limits throughput and latency.
To improve overall \Soteria throughout at a guaranteed latency, we can configure 
a $32$-node system with an admission control that limits the input rate to be
under $450$ transactions/s.  This configuration can support throughput up 
to $450$ transactions/s 
with $10$-second latency.  When the overall throughput 
requirement is higher than $450$, \Soteria
can configure another $32$-node system to satisfy throughput scalability at
the same latency level.

\section{Conclusion}
\label{sec:con}

This paper presented \Soteria, consisting of a user right management
system (URM), a distributed ledger (DLT), and a auditing service (ATS)
to support provable, auditable and scalable data governance.
To protect consumer rights of privacy and to comply with data-privacy
regulations, we design \Soteria to fulfill the functional requirements
of consent, record keeping, and transparency (for auditing).
We presented the architecture of \Soteria, its functional specifications, and
protocol choices for its base chain and side chains.  
On protocol selection, we qualitatively evaluated four 
algorithms and experimented with two readily deployable  
protocols Tendermint and Stellar, and select Tendermint
to be our side-chain protocol because of its scalability in throughput and latency.  

We have deployed \Soteria at a hospital chain, and experiment with
a news aggregator to facilitate them being CCPA-compliant. As the user base is
expected to grow, our future work will address various performance optimization issues 
that will arise with URM and DLT.  Three specific topics are:
\begin{itemize}
    \item SQL optimization. In Section~\ref{sec:esa-enforcement}, we note that the SQL query, while correct, might not be very efficient, as it includes at least one clause for each owner of the data in the database. In our experience, this is sufficient, as in practice the same sharing agreement is used by many different consumers, so clauses can be unified when the query is constructed. As consumers' preferences become more fine-grained, this might not be the case. Future work should investigate an optimization algorithm or an index structure suitable to queries of this form, so that both performance and formal correctness can be maintained.
    \item Inter-chain protocol optimization. How often side chains should hash a block onto the main chain affects \Soteria's latency and throughput.  While \Soteria may have full control on the parameters
    of its side chains, the main, public chain (such as Ethereum classic) can be shared with other applications. \Soteria should look into dynamic adjustment policies on its side-chain block size and
    hash (to the main chain) frequency.
    \item Interoperability with native access control policies. Most companies have had
    their own access control policies and/or tools.  For instance, Zelkova~\cite{8602994} developed 
    by AWS is an
    SMT-based (Satisfiability Modulo Theories) reasoning tool for analyzing security/privacy policies 
    and their risks. We will investigate how \Soteria can complement existing tools. 
\end{itemize}

\bibliographystyle{abbrv}
\bibliography{acmart}
\end{document}